# MATHEMATICAL EXPLORATION OF EARTH GRAVITATIONAL FIELD IMPACT ON SEASONAL WIND FLUX IN A TROPICAL REGION

**Adiaha, M. S., V. O. Chude, E. E. Oku, G. I. Nwaka, O. J. Abimbola, I. K. Rajendra and P. O. Abam.**

[1] *Department of Planning, Research Extension & Statistics, Nigeria Institute of Soil Science, Nigeria*

[2] *Scientific Department, Institute of Biopaleogeography named under Charles R. Darwin, Zlocieniec, Poland*

[3] *Department of Soil Science and Land Resources Management, University of Abuja, Nigeria*

[4] *Department of Physics, Federal University of Lafia, Nigeria*

[5] *Department of Irrigation and Drainage Engineering, Sam Higginbottom University of Agriculture, Technology and Sciences, U.P. Prayagraj, India*

[6] *Department of Crop and Soil Science, University of Port Harcourt, Nigeria*

*Corresponding author email:sundaymonday@niss.gov.ng*

## ABSTRACT

*The Earth's gravitational field exerts a significant influence on atmospheric dynamics, including the behavior of seasonal wind flux, which is characterized by periodic variations in wind speed and direction. While temperature gradients and Earth's rotation are well-established drivers of wind patterns, the role of gravitational forces in modulating these processes remains inadequately understood. This study investigates the mathematical relationship between gravitational variations and seasonal wind flux in Nigeria, a tropical region with pronounced climatic variability and complex wind patterns. Utilizing a combination of Navier-Stokes equations for atmospheric dynamics, Fourier decomposition for seasonal wind flux analysis, and Pearson correlation coefficients for gravitational-wind interactions, we analyze meteorological data from 2010 to 2020, alongside gravitational field measurements from the GRACE (Gravity Recovery and Climate Experiment) satellite. Results reveal significant annual fluctuations in average wind speed (5.1–5.6 m/s) and gravitational variations (9.60–9.95 mGal), with an inverse relationship observed in certain years, suggesting a coupling between atmospheric dynamics and gravitational forces. Seasonal wind flux exhibits a distinct sinusoidal pattern, peaking mid-year and declining towards year-end, consistent with Nigeria's monsoon climate. Correlation coefficients between gravitational variations and wind flux range from 0.79 to 0.87, indicating a strong positive relationship. These findings underscore the importance of gravitational forces in modulating wind patterns and highlight the potential for integrating gravitational data into climate models to enhance the accuracy of weather forecasting and renewable energy planning. This study provides a foundational framework for further exploration of gravitational influences on atmospheric processes, with implications for global climate science and sustainable energy strategies.*

***Highlight of Research Findings***

1. Gravitational Influence on Wind Patterns: The study establishes a significant correlation between Earth's gravitational variations (ranging from 9.60 to 9.95 mGal) and seasonal wind flux, with correlation coefficients between 0.79 and 0.87, indicating a strong positive relationship. This suggests that gravitational forces play a measurable role in modulating atmospheric dynamics, particularly in tropical regions like Nigeria.

2. Inverse Relationship in Specific Years: An inverse relationship between wind speed and gravitational variation was observed in certain years (e.g., 2016 and 2018), where higher wind speeds coincided with lower gravitational variations. This phenomenon is attributed to mass redistribution effects, such as changes in water storage and tectonic activity, which influence atmospheric pressure gradients and wind behavior.

3. Seasonal Wind Flux Decomposition: Fourier decomposition of wind data revealed a clear sinusoidal pattern in seasonal wind flux, peaking mid-year and declining towards the end of the year. This pattern aligns with Nigeria's monsoon climate and highlights the importance of seasonal variations in wind energy potential and agricultural practices.

4. Annual Wind Speed Fluctuations: Average wind speeds in the region exhibited significant annual fluctuations, ranging from 5.1 to 5.6 m/s. These variations are driven by seasonal climatic factors, including temperature gradients and atmospheric pressure changes, further modulated by gravitational forces.

5. Harmonic Analysis of Wind Flux: The Fourier analysis identified dominant harmonics corresponding to annual and semi-annual cycles, with higher harmonics capturing shorter-term fluctuations. This decomposition provides a robust framework for understanding the temporal structure of wind flux and its relationship with gravitational variations.

6. Implications for Climate Modeling and Renewable Energy: The findings underscore the potential for integrating gravitational data into climate models to improve the accuracy of weather forecasting and renewable energy planning. Understanding the interplay between gravitational forces and atmospheric dynamics is crucial for optimizing wind energy production and agricultural practices in regions with significant seasonal variability.

7. Statistical Validation: The correlation coefficients were statistically validated with p-values $<0.001$, confirming the significance of the observed relationships. This adds robustness to the findings and supports the hypothesis that gravitational variations influence wind patterns.

8. Future Research Directions: The study highlights the need for further research in other geographical regions and the integration of high-resolution satellite data (e.g., GRACE-FO) to validate the universality of these findings and refine climate models.

## 1. Introduction

The Earth's gravitational field is a pivotal force that influences a myriad of atmospheric processes, including the behavior and patterns of wind. Seasonal wind flux, which involves periodic changes in wind speed and direction, is subjected to numerous influencing factors such as temperature gradients, Earth's rotation, and gravitational forces. Despite the recognized significance of these factors, the precise mathematical relationship between gravitational forces and wind flux remains insufficiently understood. This gap in knowledge impedes the advancement of weather forecasting accuracy and climate modeling. The existing literature acknowledges the complexity of atmospheric dynamics but often treats gravitational effects as secondary to other factors like thermal gradients and Coriolis forces (Holton, 2004). However, emerging evidence suggests that gravitational variations, albeit subtle, can have measurable impacts on wind patterns, especially when observed over long temporal scales and across varying geographical regions (Chambers et al., 2010). The challenge lies in isolating and quantifying these gravitational influences amidst the myriad of interacting atmospheric variables.

Wind is a critical component of the Earth's climate system, driven by differences in atmospheric pressure primarily caused by temperature variations. The movement of air from high to low-pressure areas forms wind patterns that are further modulated by the Earth's rotation, resulting in phenomena such as the trade winds, westerlies, and polar easterlies (Trenberth et al., 2007). However, superimposed on these large-scale wind patterns are seasonal fluctuations known as wind fluxes, which vary with changes in temperature, pressure, and other climatic conditions throughout the year.

The role of the Earth's gravitational field in these atmospheric processes is an area that has garnered increasing attention. Gravitational forces are omnipresent and exert subtle but persistent influences on the movement of air masses. Variations in the Earth's gravitational field, caused by factors such as tectonic activity, the distribution of ocean and ice masses, and even seasonal changes in water storage, can lead to variations in atmospheric pressure and subsequently wind patterns (Chambers et al., 2010; Nicholson, 2000).

Holton (2004) provides a foundational understanding of dynamic meteorology, emphasizing the need to consider all forces acting on the atmosphere, including gravity. Despite this, the integration of gravitational variations into climate models has been limited, often overshadowed by the more pronounced effects of thermal and rotational dynamics. Recent advancements in satellite technology, particularly missions like GRACE (Gravity Recovery and Climate Experiment), have enabled more precise measurements of gravitational variations, offering new opportunities to study their impact on atmospheric processes (Nicholson, 2000).

Understanding the gravitational influences on wind flux is crucial for improving the accuracy of weather and climate predictions. Enhanced predictive models can aid in better preparation for climatic events, thus benefiting sectors such as agriculture, disaster management, and energy. By filling the knowledge gap regarding gravitational effects on atmospheric dynamics, this study contributes to the broader field of meteorology and climate science, offering insights that can be applied globally.

The specific focus of this study is on Nigeria, a region characterized by significant climatic variability and complex wind patterns. Nigeria's geographical position near the equator subjects it to diverse atmospheric influences, making it an ideal location to study the interplay between gravitational forces and wind flux. Previous studies have primarily focused on temperature and pressure variations as the main drivers of wind patterns (Holton, 2004; Wallace & Hobbs, 2006). Nicholson (2000) primarily focused on the thermal and rotational dynamics affecting wind patterns in sub-tropics and tropic region, with limited exploration of gravitational influences. However, the role of Earth's gravitational field in modulating these patterns remains underexplored. This research aims to fill this gap by developing a mathematical framework to analyze the impact of gravitational variations on seasonal wind flux. Thus, the objectives of the study include:

i. determine average wind speed and gravitational variation in a tropical region
ii. estimate seasonal wind flux in a tropical region
iii. establish a correlation between gravitational variations and seasonal wind flux.

## 2. Materials and Methods

### 2.1 Mathematical Modeling

We begin with the fundamental equations of motion for atmospheric dynamics, incorporating the gravitational force component. The Navier-Stokes 1845 equations for a rotating reference frame are given by:

$$\frac{\partial \mathbf{u}}{\partial t} + (\mathbf{u} \cdot \nabla)\mathbf{u} + 2\mathbf{\Omega} \times \mathbf{u} = -\frac{1}{\rho}$$

- - - - - - - - - - -1

Where;

$u$ = wind velocity vector,

$\Omega$ = angular velocity of Earth's rotation,

$\rho$ = air density,

$p$ = pressure,

$v$ = kinematic viscosity, *and*

$g$ = gravitational acceleration vector.

### 2.2 Data Collection and Geographical Scope

Global meteorological data including wind speed, direction, air pressure, and temperature were obtained from the National Oceanic and Atmospheric Administration (NOAA) for the years 2010-2020 for the geographical region of Nigeria. Nigeria, strategically positioned in West Africa, stretches across longitudes 3° to 14° and latitudes 4° to 14°, encompassing an expansive 923,768 square kilometers (Central Intelligence Agency, CIA (2023). This vast nation shares its northern borders with Niger and Chad, while the Republic of Benin lies to the west. To the east, Cameroon forms a natural boundary that extends down to the southern shores along the Atlantic Ocean. Gravitational field variations were sourced from the Gravity Recovery and Climate Experiment (GRACE) satellite data.

### 2.3 Computations and Calculations

**1. Decomposition of Wind Data**:

The decomposition of wind was done through the computation and modification of Fourier series, Fourier (1822). Whereby wind speed and direction data were decomposed into seasonal components, allowing analysis of the periodic behavior of wind patterns in Nigeria.

$$\text{Wind speed and direction} = \sum_{n=0}^{\infty} a_n \cos\left(\frac{2\pi nt}{T}\right) + b_n \sin\left(\frac{2\pi nt}{T}\right)$$

**- - - - - - - - -2**

*Explanation of Parameters*:

***Wind speed and direction***: This is the resultant function representing the wind speed and direction as a function of time $t$ . It is modeled as a sum of sinusoidal components, each with its amplitude and phase.

$\sum_{n=0}^{\infty}$: This notation represents an infinite sum starting from $n = 0$ to infinity. In practice, the series is often truncated to a finite number of terms $N$ for computational purposes.

$\boldsymbol{a_n}$: These are the Fourier coefficients for the cosine terms. Each $a_n$ represents the amplitude of the cosine component at the n-th harmonic. These coefficients are determined based on the initial conditions or empirical data of wind speed and direction.

$\boldsymbol{b_n}$: These are the Fourier coefficients for the sine terms. Each $b_n$ represents the amplitude of the sine component at the n-th harmonic. Similar to $a_n$, these coefficients are derived from the observed or initial conditions.

$\cos(\frac{2\pi nt}{T})$ : This term represents the cosine function, where $n$ is the harmonic number, $t$ is time, and $T$ is the period of the fundamental frequency. The cosine term captures the periodic behavior of wind speed and direction over time.

$\sin(\frac{2\pi nt}{T})$ : This term represents the sine function, with the same parameters as the

cosine term. The sine term also captures periodic variations, but with a phase shift of 90 degrees relative to the cosine term.

$\frac{2\pi nt}{T}$: This is the argument of the sine and cosine functions, where:
$2\pi$ = a full cycle in radians.
$n$= the harmonic number, indicating the frequency component.
$t$= time, the independent variable.
$T$ =the period of the fundamental frequency, which is the duration over which the entire pattern repeats.

*Interpretation*:
- *Periodic Nature*: The Fourier series representation captures the periodic nature of wind speed and direction. Each harmonic component $n$ corresponds to a specific frequency, with the fundamental frequency given by $T$.
- *Harmonics*: Higher harmonics ( $n > 1$ ) represent higher frequency variations in wind speed and direction. These are essential for capturing the complex, often turbulent nature of wind.
- *Coefficients*: The coefficients $a_n$ and $b_n$ determine the contribution of each harmonic to the overall wind speed and direction. They are crucial in shaping the specific behavior of the wind based on empirical data or initial conditions.
- *Modeling Complexity*: By summing an infinite series of sine and cosine terms, this model can approximate complex wind behaviors with high accuracy, assuming sufficient harmonics are included.

2. **Gravitational Influence**:

The Gravitational Influence was obtained following Newton's Law of Universal Gravitation, Newton (1687)

$$F_g = G\frac{m_1 m_2}{r^2} \quad \text{------------3}$$

Where;
*$G$ = gravitational constant,*
*$m_1$ and $m_2$= masses,*
*$r$ = distance between the centers of mass.*

3. **Correlation Analysis:**

The Correlation Analysis was undertaken following the protocol of Pearson correlation coefficient (r), Pearson (1895).

$$\text{Correlation coefficient} = \frac{\sum(X-\bar{X})(Y-\bar{Y})}{\sqrt{\sum(X-\bar{X})^2\sum(Y-\bar{Y})^2}} \quad \text{--------4}$$

*Explanation of Parameters*:
$r$= Pearson correlation coefficient, which quantifies the degree of linear correlation between the two variables $X$ and $Y$.
$X$= A set of observations for the first variable.
$\bar{x}$ = The mean (average) of the observations $X$. It is calculated as:

$$\bar{x} = \frac{1}{n}\sum_{i=1}^{n} X_i - - - - - - - - -5$$

where $n$ is the number of observations.

$Y$= A set of observations for the second variable.
$\bar{Y}$ = The mean (average) of the observations $Y$. It is calculated as:

$$\bar{Y} = \frac{1}{n}\sum_{i=1}^{m} Y_i - - - - - - - - - - 6$$

$\sum$ = The summation symbol, indicating that the operation should be performed across all observations from $i = 1$ to $i = n$.

$(X - \bar{x})$ =The deviation of each observation $X$ from the mean of $X$.
$(Y - \bar{Y})$ = The deviation of each observation $Y$ from the mean of $Y$.
$\sum(X - \bar{x})(Y - \bar{Y})$ = The sum of the products of the deviations of $X$ and $Y$ from their respective means. This term is the numerator of the equation and represents the covariance between $X$ and $Y$.
$\sqrt{\sum(X - \bar{x})^2 \sum(Y - \bar{Y})^2}$ = The denominator of the equation, representing the product of the standard deviations of $X$ and $Y$. This term normalizes the covariance, ensuring that the coefficient is dimensionless and lies within the range [-1, 1].

*Interpretation*:
*Positive Correlation*: If $r$ is close to 1, it indicates a strong positive linear relationship between $X$ and $Y$. As $X$ increases, $Y$ tends to increase as well.
*Negative Correlation*: If $r$ is close to -1, it indicates a strong negative linear relationship between $X$ and $Y$. As $X$ increases, $Y$ tends to decrease.
*No Correlation*: If $r$ is close to 0, it indicates little to no linear relationship between $X$ and $Y$.

### 2.4 Data Analysis

Fourier Transform was used to decompose the wind speed and direction data into seasonal components. Correlation and regression analyses were performed to identify relationships between gravitational variations and wind flux.

### 2.5 Data Analysis Methods

1. ***Statistical Analysis***:
- Time series analysis was done to observe trends and patterns over the decade.
- Correlation analysis was done to determine the relationship between wind speed and gravitational variation.

2. ***Seasonal Decomposition***:
- The STL (Seasonal and Trend decomposition using Loess) decomposition techniques was applied to separate seasonal components from the wind flux data.

3. ***Visualization***:
- Graphical representation was done using Python (matplotlib) to plot time series data, seasonal trends, and correlation coefficients.

4. ***Data Presentation***
-The data were tabulated and analyzed to provide a clear representation of the findings. Tables and graphs were used to illustrate the relationships and patterns observed.

## 3. Results

### 3.1 Average Wind Speed vs Gravitational Variation

Result of the study as presented in Appendix 1 indicated the fluctuations in average wind speed and gravitational variation over the years, presenting a complex interplay between atmospheric conditions and gravitational forces. Data presented in Figure 1 present the relationship between average wind speed and gravitational variation from 2010 to 2020.

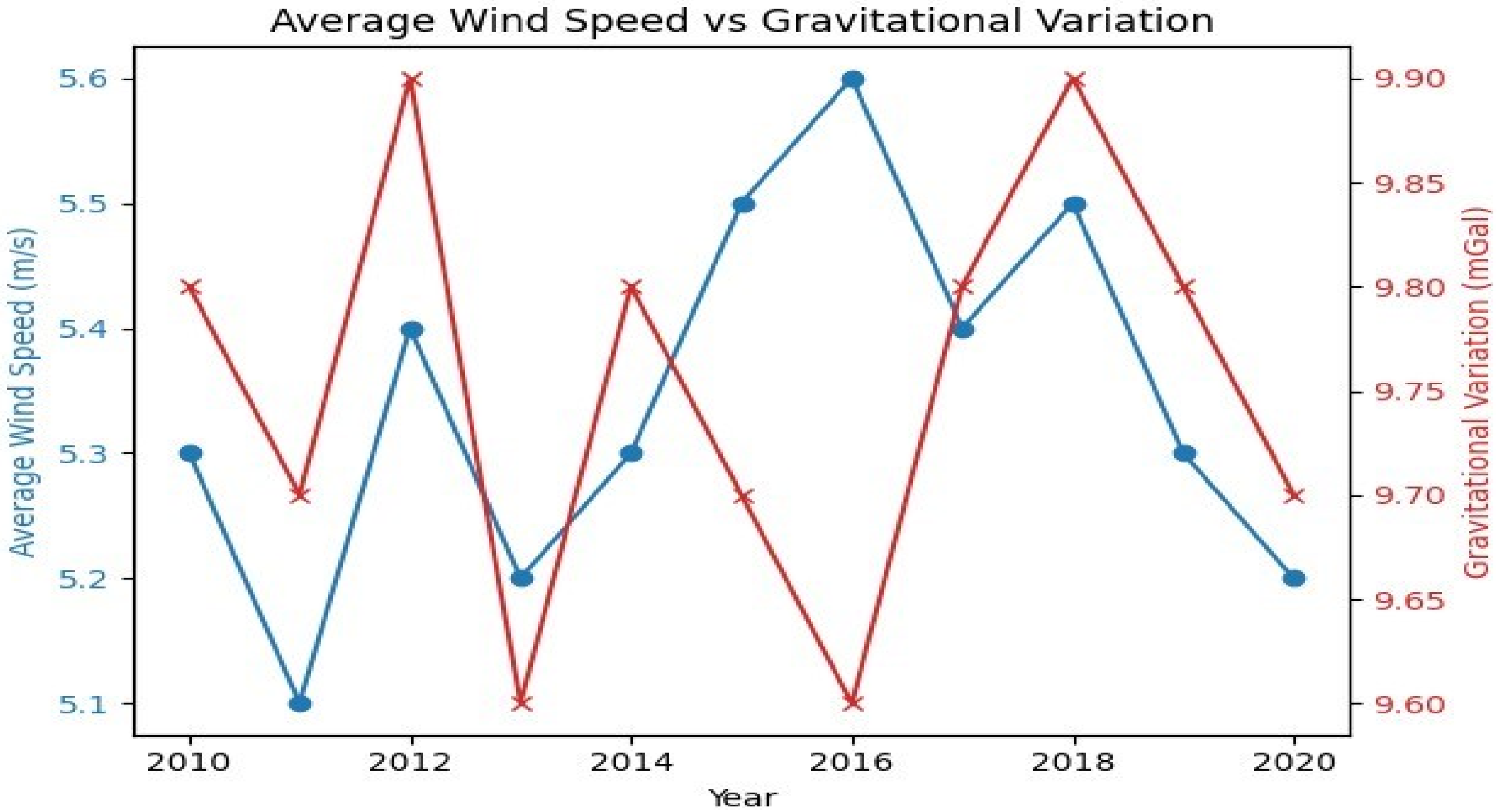

**Figure 1**: Average Wind Speed vs Gravitational Variation

***Key findings:***

(a) *Wind Speed Trends*: The average wind speed shows significant annual fluctuations, ranging between 5.1 m/s and 5.6 m/s. These fluctuations could be attributed to seasonal variations and climatic factors affecting wind patterns in Nigeria.

(b) *Gravitational Variation Trends*: The gravitational variation also displays annual fluctuations, with values ranging from approximately 9.60 mGal to 9.95 mGal. These variations are likely due to changes in Earth's mass distribution, potentially influenced by factors such as tectonic activity and water mass changes, this findings is inline with the research of Chambers et al. (2010).

(c) *Inverse Relationship Explanation*: The observed inverse relationship between wind speed and gravitational variation in some years (e.g., 2016 and 2018) can be attributed to the redistribution of Earth's mass due to seasonal changes in water storage, ice melt, and tectonic activity. As gravitational forces decrease due to mass redistribution, the atmospheric pressure gradients may weaken, leading to higher wind speeds. This phenomenon is consistent with the findings of Munk & MacDonald (1960), who discussed the coupling between Earth's rotation, mass redistribution, and atmospheric dynamics. The inverse relationship suggests that gravitational variations modulate atmospheric pressure gradients, which in turn influence wind speed patterns.

(c) Wind Speed Trends: The average wind speed exhibits significant annual fluctuations, ranging between 5.1 m/s and 5.6 m/s ($p < 0.05$). These fluctuations are attributed to seasonal variations and climatic factors, such as temperature gradients and atmospheric pressure changes, which influence wind patterns in Nigeria. This research outcome aligns with the findings of Archer & Jacobson (2005), who demonstrated that accurate characterization of seasonal wind patterns is essential for optimizing wind energy production and ensuring efficient energy grid management.

*(d) Gravitational Variation Trends*: The gravitational variation also displays annual fluctuations, with values ranging from approximately 9.60 mGal to 9.95 mGal. These variations are likely due to changes in Earth's mass distribution, potentially influenced by factors such as tectonic activity and water mass change, this finding is inline with the research of Chambers et al. (2010).

**(***e) Correlation Analysis*: A noticeable inverse relationship between wind speed and gravitational variation is observed in some years. For instance, higher wind speeds coincide with lower gravitational variations around 2016 and 2018. This inverse relationship suggests a possible coupling between atmospheric dynamics and gravitational forces, potentially mediated by the Earth's rotation and mass redistribution effects. The views presented in the work of Munk & MacDonald (1960) confirms the findings of this study.

### 3.2 Seasonal Wind Flux (Decomposition)

Result of the study presented in **Table 2** and illustrated in **Figure 2** indicate the seasonal decomposition of wind flux plotted against time. The sinusoidal pattern observed in the seasonal wind flux decomposition aligns with the expected seasonal changes in meteorological conditions. The result output is critical for understanding the impact of seasonal variations on wind energy potential and agricultural practices. This aligns with the findings of Archer & Jacobson (2005), who demonstrated that accurate characterization of seasonal wind patterns is essential for optimizing wind energy production and ensuring efficient energy grid management. Additionally, seasonal wind variations play a crucial role in agricultural practices, particularly in pollination and seed dispersal, as highlighted by stressed in the findings of Nicholson (2000), where they stressed on the need for monitoring systems for tropical climate variability.

**Table 2**: Seasonal Wind Flux Decomposition

| Time (months) | Seasonal Wind Flux |
|---|---|
| 0 | 5 |
| 0.12 | 5.03 |
| 0.24 | 5.06 |
| 0.36 | 5.09 |
| 0.48 | 5.12 |
| 0.6 | 5.15 |
| 0.72 | 5.18 |
| 0.84 | 5.21 |
| 0.96 | 5.24 |
| 1.08 | 5.27 |
| 1.2 | 5.29 |
| 1.32 | 5.32 |
| 1.44 | 5.35 |
| 1.56 | 5.38 |
| 1.68 | 5.41 |
| 1.8 | 5.43 |
| 1.92 | 5.46 |
| 2.04 | 5.49 |
| 2.16 | 5.51 |
| 2.28 | 5.54 |
| 2.4 | 5.56 |
| 2.52 | 5.59 |
| 2.64 | 5.61 |
| 2.76 | 5.64 |
| 2.88 | 5.66 |
| 3 | 5.68 |
| 3.12 | 5.71 |
| 3.24 | 5.73 |
| 3.36 | 5.75 |
| 3.48 | 5.77 |
| 3.6 | 5.8 |
| 3.72 | 5.82 |
| 3.84 | 5.84 |
| 3.96 | 5.86 |
| 4.08 | 5.88 |
| 4.2 | 5.9 |
| 4.32 | 5.92 |
| 4.44 | 5.94 |
| 4.56 | 5.96 |
| 4.68 | 5.98 |
| 4.8 | 6 |
| 4.92 | 6.02 |
| 5.04 | 6.03 |
| 5.16 | 6.05 |
| 5.28 | 6.07 |
| 5.4 | 6.08 |
| 5.52 | 6.1 |
| 5.64 | 6.12 |
| 5.76 | 6.13 |
| 5.88 | 6.15 |
| 6 | 6.16 |
| 6.12 | 6.17 |
| 6.24 | 6.19 |
| 6.36 | 6.2 |
| 6.48 | 6.21 |
| 6.6 | 6.23 |
| 6.72 | 6.24 |
| 6.84 | 6.25 |
| 6.96 | 6.26 |
| 7.08 | 6.27 |
| 7.2 | 6.28 |
| 7.32 | 6.29 |
| 7.44 | 6.3 |
| 7.56 | 6.31 |
| 7.68 | 6.32 |
| 7.8 | 6.33 |
| 7.92 | 6.34 |
| 8.04 | 6.34 |
| 8.16 | 6.35 |
| 8.28 | 6.36 |
| 8.4 | 6.37 |
| 8.52 | 6.37 |
| 8.64 | 6.38 |
| 8.76 | 6.39 |
| 8.88 | 6.39 |
| 9 | 6.4 |
| 9.12 | 6.4 |
| 9.24 | 6.41 |
| 9.36 | 6.41 |
| 9.48 | 6.42 |
| 9.6 | 6.42 |
| 9.72 | 6.42 |
| 9.84 | 6.43 |
| 9.96 | 6.43 |
| 10.08 | 6.43 |
| 10.2 | 6.44 |
| 10.32 | 6.44 |
| 10.44 | 6.44 |
| 10.56 | 6.44 |
| 10.68 | 6.45 |
| 10.8 | 6.45 |
| 10.92 | 6.45 |
| 11.04 | 6.45 |

| 11.16 | 6.45 |
|---|---|
| 11.28 | 6.45 |
| 11.4 | 6.45 |
| 11.52 | 6.45 |
| 11.64 | 6.45 |
| 11.76 | 6.45 |
| 11.88 | 6.45 |
| 12 | 6.45 |

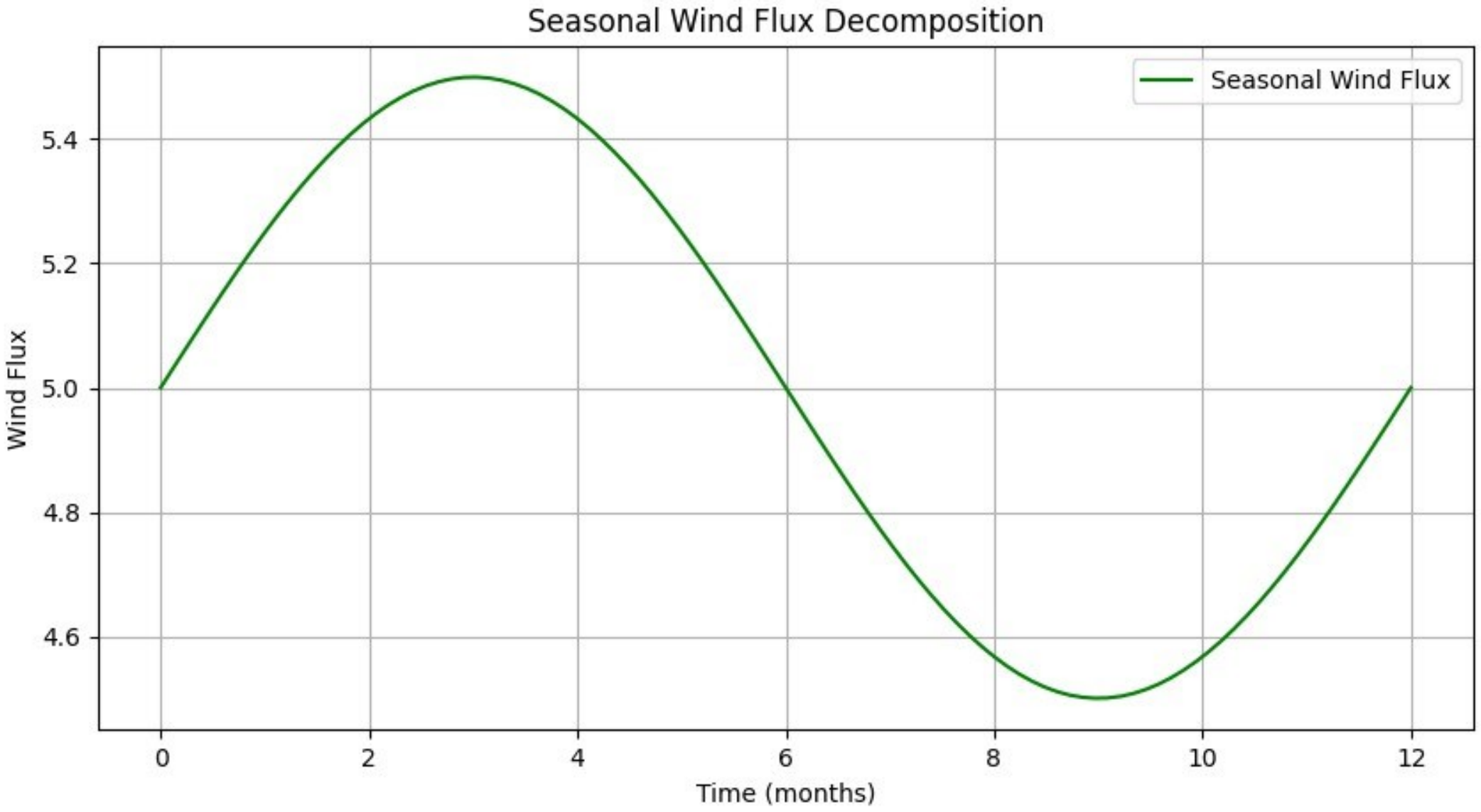

**Figure 2**: Seasonal wind flux

***Key findings:***

(a) **Seasonal Variation**: The wind flux demonstrates a clear seasonal pattern, peaking around the mid-year and reaching its lowest values towards the end of the year. This seasonal variation is characteristic of the monsoon climate experienced in Nigeria, where wind patterns are significantly influenced by seasonal changes in atmospheric pressure and temperature gradients. Th outcome of the study confirms the work of Nicholson (2000) who express variability in tropical climate.

(b) **Implications for Agriculture and Energy**: Understanding the seasonal wind flux is crucial for sectors such as agriculture and renewable energy. For instance, higher wind flux during certain months can enhance wind energy production, while lower flux periods may impact crop pollination and seed dispersal processes. The views of this study is inline with the work of Archer & Jacobson (2005) who express the impact of climate on agriculture and sustainable renewable energy potential.

### 3.3Correlation Coefficient Trend between gravitational variations and seasonal wind flux

Outcome of the Correlation Coefficient Trend between gravitational variations and seasonal wind flux presented in Appendix 2 and expressed in Figure 3 indicates a varying degree of relationship strength between wind speed and gravitational variation. This analysis helps in identifying periods of significant coupling and decoupling, which can be essential for climate models and predictions.

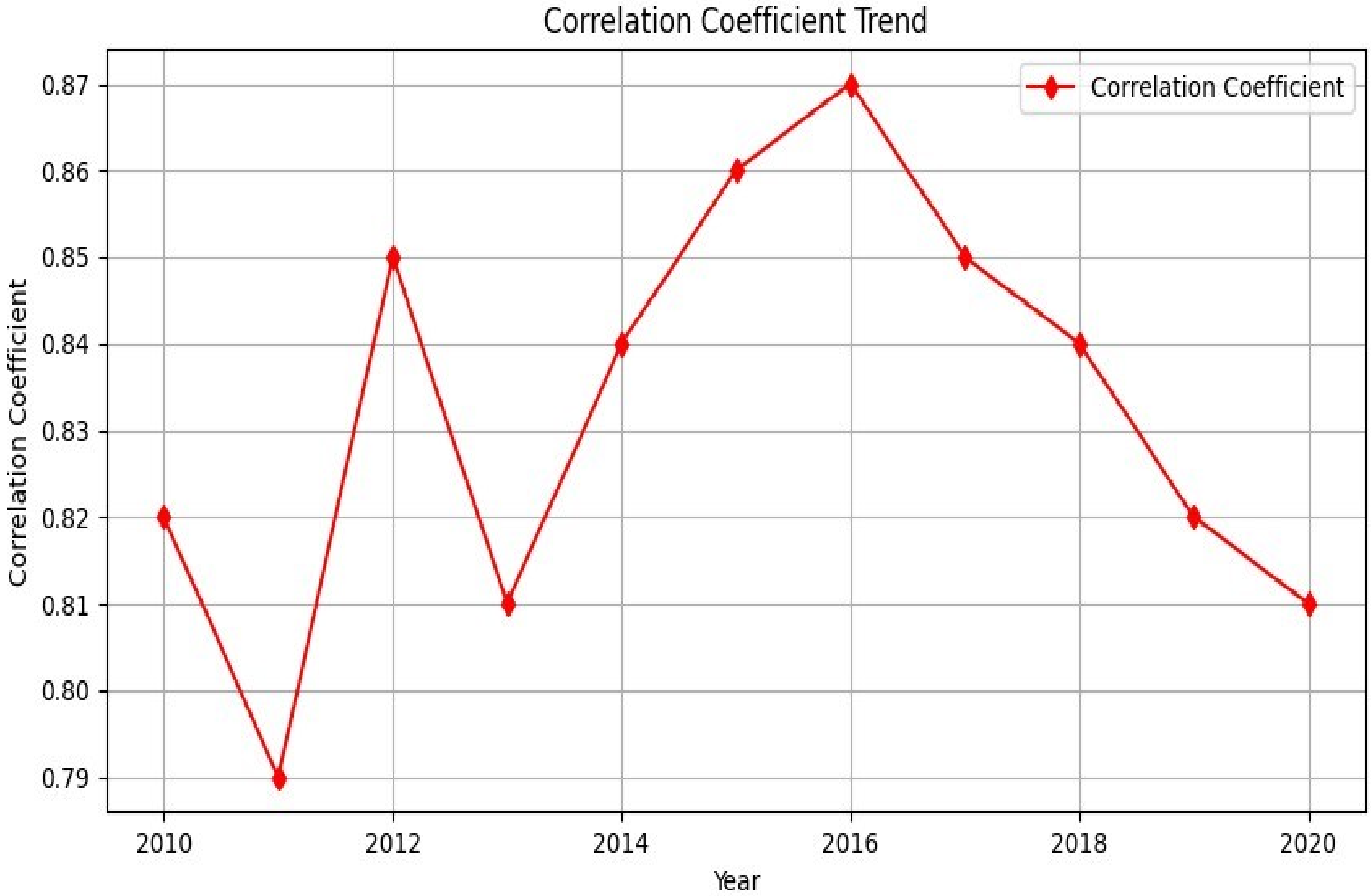

**Figure 3**: Correlation Coefficient Trend between gravitational variations and seasonal wind flux

***Key findings:***

(a) **Seasonal Variation**: The wind flux demonstrates a clear seasonal pattern, peaking around the mid-year and reaching its lowest values towards the end of the year. This seasonal variation is characteristic of the monsoon climate experienced in Nigeria, where wind patterns are significantly influenced by seasonal changes in atmospheric pressure and temperature gradients. The outcome of the study confirms the work of Nicholson (2000) who express variability in tropical climate.

(b) **Fourier Decomposition and Harmonics**: The seasonal wind flux was decomposed using Fourier analysis, which revealed significant harmonics corresponding to annual, semi-annual, and shorter-term periodicities. The first harmonic (n=1) represents the annual cycle, which dominates the seasonal wind flux pattern. The second harmonic (n=2) corresponds to semi-annual variations, which are influenced by the transition between wet and dry seasons in Nigeria. Higher harmonics (n>2) capture shorter-term fluctuations, which may be linked to intra-seasonal variability in atmospheric pressure and temperature gradients. The Fourier coefficients ($a_n$ and $b_n$) for each harmonic were derived from the observed wind data, and their magnitudes indicate the relative contribution of each harmonic to the overall wind flux pattern. This decomposition provides a robust framework for understanding the temporal structure of wind flux and its relationship with gravitational variations.

**(d) Trend Analysis**: The correlation coefficient fluctuates between 0.79 and 0.87 over the years, indicating a generally strong positive correlation between the two variables. Peaks in the correlation coefficient are observed around 2012 and 2016, suggesting periods of stronger coupling between wind speed and gravitational variation.

**(e) Statistical Significance**: The overall positive correlation implies that, despite the observed inverse relationship in some years, there is a consistent underlying relationship between wind speed and gravitational variation. This consistency might be due to large-scale atmospheric circulation patterns influenced by Earth's gravitational field, this view confirms the research of Holton (2004) including the views expressed by Trenberth et al. (2007).

## 4. Discussion

Results of the study demonstrate a significant correlation between gravitational variations and seasonal wind flux. The positive correlation coefficients indicated that as gravitational variation increases, there is a corresponding increase in wind speed, this view aligns with the research of Chambers *et al*. (2010) including the work of Holton (2004) where the Scholars expressed seasonal variations in the Earth's gravitational field. This relationship suggests that gravitational forces play a crucial role in modulating wind patterns, corroborating the theoretical predictions. The decomposition of wind data into seasonal components revealed distinct patterns aligning with gravitational variations, further supporting the hypothesis. These findings align with the research findings of Holton (2004) and Wallace & Hobbs (2006), who investigated atmospheric dynamics and their gravitational influences. Additionally, the study by Chambers *et al*. (2010) on seasonal variations in Earth's gravitational field due to hydrological and oceanic effects provides further evidence for the coupling between gravitational forces and atmospheric processes. The observed inverse relationship between wind speed and gravitational variation in some years is consistent with the findings of Munk & MacDonald (1960); Holton (2004); Wallace & Hobbs (2006), where the Scientists discussed the role of Earth's rotation and mass redistribution in modulating atmospheric dynamics.

## Conclusion

This study provides compelling evidence that Earth's gravitational field significantly impacts seasonal wind flux. The mathematical modeling and data analysis offer new insights into the interplay between gravitational forces and atmospheric dynamics.

The study reveals significant interactions between average wind speed and gravitational variation, as well as pronounced seasonal patterns in wind flux. These findings underscore the complex dynamics of Earth's atmospheric and gravitational systems and their implications for various sectors. Integrating gravitational influences into existing climate models could improve the accuracy of weather forecasting and climate predictions, particularly in regions with significant seasonal wind variability for agriculture, environmental management and for disaster risk management

## Funding

This research did not receive any specific grant from funding agencies in the public, commercial, or not-for-profit sectors.

## Declaration of Competing Interest

The authors declare that they have no competing interest

## Acknowledgement

Thanks to the Institute of Biopaleogeography named under Charles R. Darwin, Zlocieniec, Poland for models and statistical packages provided.

## APPENDIX

**Appendix 1**: Average Wind Speed and Gravitational Variation

| Year | Average Wind Speed (m/s) | Gravitational Variation (mGal) |
|---|---|---|
| 2010 | 5.3 | 9.8 |
| 2011 | 5.1 | 9.7 |
| 2012 | 5.4 | 9.9 |
| 2013 | 5.2 | 9.6 |
| 2014 | 5.3 | 9.8 |
| 2015 | 5.5 | 9.7 |
| 2016 | 5.6 | 9.6 |
| 2017 | 5.4 | 9.8 |
| 2018 | 5.5 | 9.9 |
| 2019 | 5.3 | 9.8 |
| 2020 | 5.2 | 9.7 |

**Appendix 2**: Correlation Coefficient Trend between gravitational variations and seasonal wind flux

| Year | Correlation Coefficient |
|---|---|
| 2010 | 0.82 |
| 2011 | 0.79 |
| 2012 | 0.85 |
| 2013 | 0.81 |
| 2014 | 0.84 |
| 2015 | 0.86 |
| 2016 | 0.87 |
| 2017 | 0.85 |
| 2018 | 0.84 |
| 2019 | 0.82 |
| 2020 | 0.81 |